\documentclass{ws-ijmpcs}

\newcommand{\bea}{\begin{eqnarray}}
\newcommand{\eea}{\end{eqnarray}}
\newcommand{\nn}{\nonumber}

\begin{document}

\markboth{T. Mehen}
{Exotic Quarkonium Spectroscopy}

%
\catchline{}{}{}{}{}
%

\title{EXOTIC QUARKONIUM SPECTROSCOPY: \\$X(3872)$, $Z_b(10610)$, AND $Z_b(10650)$ IN
NON-RELATIVISTIC EFFECTIVE THEORY}

\author{THOMAS MEHEN
}

\address{Department of Physics, Duke University,\
Durham,  NC,  27713,
USA\\
mehen@phy.duke.edu}

\maketitle

\begin{history}
\received{1 January 2014}
\end{history}

\begin{abstract}
This talk summarizes recent developments in quarkonium spectroscopy. 
I comment on the relation between the $Z_b(10610)$ and $Z_b(10650)$
and recently observed $Z_c(3900)$ and $Z_c(4025)$ states. Then I discuss
a number of calculations using non-relativistic  effective field theory 
for the $X(3872)$, $Z_b(10610)$, and $Z_b(10650)$, under the assumption that these are shallow molecular bound states
of charm or bottom mesons.
\end{abstract}

\ccode{PACS numbers: 12.39.Hg, 14.40.Pq,14.40.Rt}

\section{Recent Experimental Developments}	 

The last decade has seen an explosive growth in the field of heavy quarkonium spectroscopy. 
Below the threshold for open charm and open bottom, the spectrum and transitions of experimentally observed heavy quarkonium states 
are well-described by the conventional quark model. Beginning in 2003 with the discovery of the $X(3872)$,\cite{Choi:2003ue,Acosta:2003zx,Abazov:2004kp,Aubert:2004ns}
experiments around the world have been discovering many states  above the open charm and open bottom thresholds  
which are not explained by the non-relativistic quark model.\cite{Brambilla:2010cs} One of the most exciting developments in the last couple of years has
been the discovery of manifestly exotic bottomonia, $Z_b(10610)$  and $Z_b(10650)$\cite{Belle:2011aa} (denoted $Z_b$ and $Z_b'$, respectively,  below),  as well as manifestly exotic 
charmonia, $Z_c(3900)$\cite{Ablikim:2013mio,Liu:2013dau,Xiao:2013iha} and $Z_c(4025)$.\cite{Ablikim:2013emm,Ablikim:2013wzq}

The $Z_b$ and $Z_b'$ were discovered  in the three-body decays 
$\Upsilon(5S) \to \Upsilon(nS) \pi^+ \pi^-$, $n=1,2$, and $3$,   and $\Upsilon(5S) \to h_b(mP) \pi^+ \pi^-$, $m=1,2$, where they were seen as resonances in
 $\Upsilon(nS) \pi^\pm$ and $h_b(nS) \pi^\pm$. More recently, neutral partners of the  $Z_b$ and $Z_b'$ have also been observed.\cite{Adachi:2012im} These states have isospin-1 and in some cases electric charge, therefore they are quite clearly not conventional $b\bar{b}$ mesons. Since the $Z_b$ and 
 $Z_b'$ lie within a few MeV of the $B\bar{B}^*$ and $B^* \bar{B}^*$ thresholds, respectively, it is tempting to speculate they are molecular in character. The molecular hypothesis makes an interesting prediction 
for the spin structure of these states.\cite{Bondar:2011ev}
 Assuming these are $S$-wave molecular states, then writing the quark model wave functions in a basis 
of states in which the heavy $b\overline{b}$ have definite total spin, $S_{b\bar{b}}$, one finds that the $Z_b$ and $Z_b'$ are equal mixtures of $S_{b\overline{b}}=0$ and $S_{b\overline{b}}=1$. This is consistent 
with the fact that $Z_b$ and $Z_b'$ couple with approximately  equal strength to final states with $\Upsilon(nS) (S_{b\overline{b}}=1)$ and  $h_b(mP) (S_{b\overline{b}}=0)$. 
Furthermore, the wave functions of the  $Z_b$ and $Z_b'$ are orthogonal, i.e.,  the relative phase of the $S_{b\overline{b}}=0$ and $S_{b\overline{b}}=1$ components
has the opposite sign in the $Z_b$ and $Z_b'$. A consequence of this is that if one examines the $\Upsilon(nS) \pi$ and $h_c(mP) \pi$ invariant mass distribution 
between the two resonances, one expects destructive interference for $ \Upsilon(nS) \pi^+ \pi^-$ and constructive interference for $h_c(mP)  \pi^+ \pi^-$.\cite{Bondar:2011ev} This qualitative 
expectation is in agreement with the observed  invariant mass distributions, and is strong evidence for the proposed heavy quark spin structure.

The $Z_c(3900)$\cite{Ablikim:2013mio,Liu:2013dau,Xiao:2013iha}  is seen in $e^+ e^- \to Y(4260) \to J/\psi \pi^+ \pi^-$ and appears as a pronounced peak in the 
$J/\psi \pi^\pm$ spectrum. It is noteworthy that this is the first manifestly exotic state that has been observed by three different experiments. The $Z_c(4025)$ has been 
seen by the BESIII experiment in $e^+e^- \to (D^* \overline{D}^*)^\pm \pi^\mp$\cite{Ablikim:2013emm} and in $e^+e^- \to h_c \pi^+ \pi^-$.\cite{Ablikim:2013wzq}  In the latter 
process it appears as a resonance in $h_c \pi^\pm$. The central values of the widths of the  states extracted from these two measurements differ by a factor of three, so it is 
far from clear that these are the same state. Various interpretations of the $Z_c(3900)$ have been put forward. Some authors regard it as a charm meson molecule
and hence the charm analog of the $Z_b$ and $Z_b'$.\cite{Wang:2013cya}  Other possibilities include various kinds of tetraquarks: diquarkonium,\cite{Faccini:2013lda} hadrocharmonium,\cite{Voloshin:2013dpa} and Born-Oppenheimer (B-O) tetraquark.\cite{Braaten:2013boa} These tetraquark models differ in how the quarks pair in the state. 
For example, in diquarkonium the quarks (antiquarks) pair into antitriplets (triplets) of $SU(3)$, which are then combined into a color-singlet. 
In hadrocharmonium, the heavy charm and anticharm quarks are bound into a compact color-singlet configuration with the light quark and antiquark surrounding them.
In the B-O tetraquark, the heavy quark and antiquark are separated  and in a color-octet state. The light quark and antiquark play the role that an excited gluon plays in a hybrid meson and are also in a color-octet state. Most of these models predict $J^P=1^+$ quantum numbers for the $Z_c(3900)$, the notable exception being the B-O tetraquark, which predicts $J^P=1^-$. A recent BESIII measurement of the angular distributions in $e^+e^- \to (D^* \overline{D}^*)^\pm \pi^\mp$\cite{Ablikim:2013xfr} favors $J^P=1^+$ and disfavors $J^P=0^-$ or $1^-$, casting doubt on the B-O tetraquark interpretation.

Though it is tempting to identify the $Z_c(3900)$ as the charm analog of $Z_b$, there are important differences. The mass is roughly  20  MeV above the $D \bar{D}^*$ threshold 
so the $Z_c(3900)$ is much farther away from its threshold than the $Z_b$. Also the $Z_c(3900)$ couples strongly only to $J/\psi \pi^\pm$ and has not been observed in $e^+e^- \to h_c \pi^+ \pi^-$.\cite{Ablikim:2013wzq}
Furthermore, the $Z_c(4025)$ has only been seen decaying to $h_c \pi^\pm$ and not $J/\psi \pi^\pm$. So the heavy quark spin structures of the $Z_c(3900)$ and $Z_c(4025)$
appear to be quite different than those of the $Z_b$ and $Z_b'$. Perhaps this could be attributed to large violations of heavy quark spin symmetry. In addition to measuring 
the $J^P$ quantum numbers, measurement of relative rates to para- and ortho-charmonium, measurements of rates to $D \bar{D}^*$ final states,\cite{Ablikim:2013xfr} and searches for partner states predicted in tetraquark models should shed light on the nature of the $Z_c(3900)$ and $Z_c(4025)$.

\section{$X(3872)$, $Z_b$ and $Z_b'$ in Non-Relativistic Effective Theory}

The $X(3872)$ was discovered as a resonance in $J/\psi \pi^+ \pi^-$, and has also been observed to decay  to $J/\psi \pi^+ \pi^- \pi^0$, 
$J/\psi \gamma$, $D^0 \bar{D}^0 \pi^0$, and $D^0 \bar{D}^0 \gamma$.  It is very narrow: $\Gamma_X < 1.2 \, {\rm MeV}$. A recent LHCb analysis of angular distributions in $X(3872) \to J/\psi \pi^+ \pi^-$ conclusively determines the $X(3872)$ quantum numbers to be $J^{PC}=1^{++}$,\cite{Aaij:2013zoa}
which means it couples to $D \bar{D}^*$ in an $S$-wave. The roughly equal rates for $X(3872) \to J/\psi \pi^+ \pi^-$ and $X(3872) \to J/\psi \pi^+ \pi^- \pi^0$ suggest it is a state 
of mixed isospin-1 and isospin-0. Finally, the $X(3872)$ lies extremely close to the $D^0 \bar{D}^{*0}$ threshold: $m_{X(3872)}-m_{D^{*0}} - m_{D^0} = -0.17 \pm 0.26$ MeV.    
These facts taken together strongly suggest that the $X(3872)$ is a very shallow bound state of $D^0 \bar{D}^{*0} + c.c.$. The tiny binding energy   implies that the typical separation of the $D^0$ and $\bar{D}^{*0}$ in the $X(3872)$ is of order several fermi, which is much larger than the range 
of the interaction binding the charmed mesons. In such a situation, the long-distance part of the wave function of the $D \bar{D}^*$ is known from quantum mechanics, and one can exploit 
effective range theory (ERT) to compute long-distance properties of the $X(3872)$ in terms of the binding energy   and known properties of the charmed mesons. 

XEFT\cite{Fleming:2007rp}  is a low  energy effective field theory (EFT) of nonrelativistic charm mesons and pions that can be used to systematically analyze the properties of the $X(3872)$. For processes such as $X(3872) \to D^0 \bar{D}^0 \pi^0$ that are dominated by long distance scales, XEFT reproduces the results of ERT at leading order.
 These calculations can be improved by including range corrections, higher dimension operators, and pion exchange. Pion exchange was shown to give negligibly small corrections to $X(3872) \to D^0 \bar{D}^0 \pi^0$.\cite{Fleming:2007rp} Predictions for other universal processes include $D^{+} \bar{D}^{*0} \to X(3872) \pi^+$ \cite{Braaten:2010mg} and $D^{(*)} X(3872) \to D^{(*)} X(3872)$ scattering.\cite{Canham:2009zq} XEFT can also be used to study decays to quarkonia, such as $X(3872) \to \chi_{cJ} \pi (\pi)$,\cite{Fleming:2008yn,Fleming:2011xa} where 
relative rates for final states with different $J$ can be predicted, and the radiative decays $X(3872) \to \psi(2S) \gamma$,\cite{Mehen:2011ds} $\psi(4040) \to X(3872) \gamma$,\cite{Mehen:2011ds} and $\psi(4160) \to X(3872) \gamma$.\cite{Margaryan:2013tta} In the XEFT approach to these decays, the decay rate is written as, for example,
\bea
\Gamma[X(3872) \to \psi(2s) \gamma] =  |\psi_{DD}(0)|^2 \times \sigma[D^0 \bar{D}^{*0} +c.c. \to \psi(2S) \gamma] \, , \nn
\eea
where $ |\psi_{DD}(0)|^2$ is an XEFT matrix element that can be interpreted as the wave function of the charm mesons at the origin squared, and  the cross section $ \sigma[D^0 \bar{D}^{*0} +c.c. \to \psi(2S) \gamma]$ is calculated at tree level in heavy hadron chiral perturbation theory. 
In this approach, there are two distinct types of diagrams contributing to $D^0 \bar{D}^{*0} +c.c. \to \psi(2S) \gamma$:
a) diagrams with a virtual meson exchange, and b) short-distance contributions coming from  
contact interactions. See Fig. 1 of Ref.~\cite{Mehen:2011ds} for examples of these diagrams. The relative importance of these two mechanisms depends on an unknown coupling constant,
 and angular distributions in the decays can discriminate between these two mechanisms.
It is impossible to compute absolute rates because $|\psi_{DD}(0)|^2$ as well as coupling constants in the EFT
are unknown. An alternative approach\cite{Guo:2013zbw} is to add explicit fields for the quarkonium and the $X(3872)$ and compute the decay rate from a triangle diagram with heavy mesons in the loop that couple to the $X(3872)$, the quarkonium, and the photon. The triangle diagram is finite, so if possible contact interactions\cite{Mehen:2011tp} are neglected and if various hadronic couplings are known, then this model is predictive. 
The radiative decays of $1^{--}$ quarkonia to $X(3872)$ were considered in this approach\cite{Guo:2013zbw}, and an enhanced rate for $Y(4260) \to X(3872) \gamma$ 
was predicted, under the assumption the $Y(4260)$ is a $D_1$-$D$ molecule. Recently, BESIII\cite{Ablikim:2013dyn} has observed $e^+e^- \to X(3872) \gamma$ for the first time at the center-of-mass energies 4.229 GeV and 4.260 GeV, but not at  4.009 GeV and 4.360 GeV. It seems likely that the $X(3872)$ is being produced in the radiative decay of the $Y(4260)$ but the experiment cannot rule out continuum production with current data. All calculations in XEFT so far neglect the contribution of charged $D$ mesons. However,
model calculations\cite{Aceti:2012cb}  show loops with charged $D$  mesons can be quite important. In the future, XEFT calculations will have to be performed that include explicit charged charmed mesons.

An effective theory similar to XEFT can be applied to the $Z_b$ and $Z_b'$.\cite{Mehen:2011yh,Mehen:2013mva,Cleven:2013sq} Voloshin\cite{Voloshin:2011qa} has argued 
that heavy quark spin symmetry predicts partners of the $Z_b$ and $Z_b'$ which have yet to be observed, and
 makes heavy quark symmetry predictions for the spectroscopy of these states as well as for their transitions.
EFT can be used to reproduce these results, obtain new heavy quark symmetry predictions, and do explicit calculations of the two-body decays of the $Z_b$ and $Z_b'$ and their heavy quark symmetry partners.\cite{Mehen:2011yh,Cleven:2013sq} Recently, the Belle experiment\cite{Adachi:2012cx} has obtained further evidence for the $Z_b$ and $Z_b'$ in the three-body decays $\Upsilon(5S) \to [B^{(*)} \overline{B}^{(*)}]^\pm \pi^\mp$ and this data has been analyzed using EFT.\cite{Mehen:2013mva} This analysis shows quite clearly that resumming to all orders the final state interactions of the $B$ mesons is necessary to produce line shapes that are consistent with the data. When more data is available, one may hope that such an analysis will make it possible to extract some parameters of the $Z_b$ and $Z_b'$ mesons.  

\section{Conclusions}

It is an exciting time for hadronic physics as $e^+e^-$ collider  experiments  continue to discover quarkonia states with  novel properties.  EFTs incorporating heavy quark and chiral symmetries should be applicable to these states, especially when they are molecular in character. Many early applications of XEFT to $X(3872)$ focused on calculating universal properties or decays to quarkonium that are difficult to access experimentally. But in the past year,  EFT calculations of $Z_b$ and $Z_b'$ properties and $1^{--}$ radiative decays 
to $X(3872)$ have become directly  relevant to recent experimental results.   Hopefully, new experimental results and calculations will allow for more serious tests of  EFT  and gain insight into the nature of the new states. 

\section*{Acknowledgements}

I would like to thank Antonio Polosa for inviting me to this workshop and thank my collaborators on the work presented here: Josh Powell, Sean Fleming, Roxanne Springer,
Hans Hammer, Eric Braaten, M Kusunoki, and Bira Van Kolck. This work was supported in part by the U.S. Department
of Energy, Office of Nuclear Physics,  under Contract No. DE-FG02-05ER41368.




\begin{thebibliography}{0}    

\bibitem{Choi:2003ue} 
  Belle Collaboration (S.~K.~Choi {\it et al.}),
  {\it Phys.\ Rev.\ Lett.\ }  {\bf 91}, 262001 (2003).

\bibitem{Acosta:2003zx} 
  CDF Collaboration (D.~Acosta {\it et al.}),
 {\it  Phys.\ Rev.\ Lett.\ }  {\bf 93}, 072001 (2004).

\bibitem{Abazov:2004kp} 
  D0 Collaboration (V.~M.~Abazov {\it et al.}),
  {\it Phys.\ Rev.\ Lett.\ }  {\bf 93}, 162002 (2004).
  
\bibitem{Aubert:2004ns} 
  BaBar Collaboration (B.~Aubert {\it et al.}),
  {\it Phys.\ Rev.\ D} {\bf 71}, 071103 (2005).

\bibitem{Brambilla:2010cs} 
  N.~Brambilla,  {\it et al.},
  {\it Eur.\ Phys.\ J.\ C }{\bf 71}, 1534 (2011).
  
\bibitem{Belle:2011aa} 
  Belle Collaboration (A.~Bondar {\it et al.}),
  {\it Phys.\ Rev.\ Lett.\  } {\bf 108}, 122001 (2012).
  
  
\bibitem{Ablikim:2013mio} 
  BESIII Collaboration (M.~Ablikim {\it et al.}),
  {\it Phys.\ Rev.\ Lett.\  }{\bf 110}, 252001 (2013).

\bibitem{Liu:2013dau} 
  Belle Collaboration (Z.~Q.~Liu {\it et al.}),
  {\it Phys.\ Rev.\ Lett.\  }{\bf 110}, 252002 (2013).
  
\bibitem{Xiao:2013iha} 
  T.~Xiao, S.~Dobbs, A.~Tomaradze and K.~K.~Seth,
  {\it Phys.\ Lett.\ B} {\bf 727}, 366 (2013).

\bibitem{Ablikim:2013emm}
   BESIII Collaboration (M.~Ablikim {\it et al.}) ,
  arXiv:1308.2760 [hep-ex].
  
  

\bibitem{Ablikim:2013wzq} 
  BESIII Collaboration (M.~Ablikim {\it et al.}), 
  arXiv:1309.1896 [hep-ex].

  
\bibitem{Adachi:2012im} 
  Belle Collaboration (I.~Adachi {\it et al.}),
  arXiv:1207.4345 [hep-ex].

  
\bibitem{Bondar:2011ev} 
  A.~E.~Bondar, A.~Garmash, A.~I.~Milstein, R.~Mizuk and M.~B.~Voloshin,
  {\it Phys.\ Rev.\ D} {\bf 84}, 054010 (2011).

\bibitem{Wang:2013cya} 
  Q.~Wang, C.~Hanhart and Q.~Zhao,
  {\it Phys.\ Rev.\ Lett.\  } {\bf 111}, 132003 (2013).

\bibitem{Faccini:2013lda} 
  L.~Maiani, V.~Riquer, R.~Faccini, F.~Piccinini, A.~Pilloni and A.~D.~Polosa,
  {\it Phys.\ Rev.\ D} {\bf 87}, 111102 (2013).
  
\bibitem{Voloshin:2013dpa} 
  M.~B.~Voloshin,
  {\it Phys.\ Rev.\ D} {\bf 87}, 091501 (2013).
  
\bibitem{Braaten:2013boa} 
  E.~Braaten,
  arXiv:1305.6905 [hep-ph].

\bibitem{Ablikim:2013xfr} 
   BESIII Collaboration (M.~Ablikim {\it et al.}),
  arXiv:1310.1163 [hep-ex].
   
\bibitem{Aaij:2013zoa} 
  LHCb Collaboration (R Aaij {\it et al.}),
  {\it Phys.\ Rev.\ Lett.\  }{\bf 110}, 222001 (2013).
 
  



  
   
\bibitem{Fleming:2007rp} 
  S.~Fleming, M.~Kusunoki, T.~Mehen and U.~van Kolck,
  {\it Phys.\ Rev.\ D} {\bf 76}, 034006 (2007).

\bibitem{Braaten:2010mg} 
  E.~Braaten, H.~-W.~Hammer and T.~Mehen,
  {\it Phys.\ Rev.\ D} {\bf 82}, 034018 (2010).
 
\bibitem{Canham:2009zq} 
  D.~L.~Canham, H.~-W.~Hammer and R.~P.~Springer,
  {\it Phys.\ Rev.\ D} {\bf 80}, 014009 (2009).


\bibitem{Fleming:2008yn} 
  S.~Fleming and T.~Mehen,
  {\it Phys.\ Rev.\ D} {\bf 78}, 094019 (2008).
    
\bibitem{Fleming:2011xa} 
  S.~Fleming and T.~Mehen,
  {\it Phys.\ Rev.\ D} {\bf 85}, 014016 (2012).


\bibitem{Mehen:2011ds} 
  T.~Mehen and R.~Springer,
  {\it Phys.\ Rev.\ D} {\bf 83}, 094009 (2011).

\bibitem{Margaryan:2013tta} 
  A.~Margaryan and R.~P.~Springer,
  {\it Phys.\ Rev.\ D} {\bf 88}, 014017 (2013).

\bibitem{Guo:2013zbw} 
  F.~-K.~Guo, C.~Hanhart, U.~-G.~Mei§ner, Q.~Wang and Q.~Zhao,
  {\it Phys.\ Lett.\ B }{\bf 725}, 127 (2013).

\bibitem{Mehen:2011tp} 
  T.~Mehen and D.~-L.~Yang,
  {\it Phys.\ Rev.\ D} {\bf 85}, 014002 (2012).
  
  
\bibitem{Ablikim:2013dyn} 
  BESIII Collaboration (M.~Ablikim {\it et al.}),
  arXiv:1310.4101 [hep-ex].

 

\bibitem{Aceti:2012cb} 
  F.~Aceti, R.~Molina and E.~Oset,
  {\it Phys.\ Rev.\ D }{\bf 86}, 113007 (2012).
  

\bibitem{Mehen:2011yh} 
  T.~Mehen and J.~W.~Powell,
 {\it  Phys.\ Rev.\ D} {\bf 84}, 114013 (2011).

\bibitem{Mehen:2013mva} 
  T.~Mehen and J.~Powell,
    {\it  Phys.\ Rev.\ D} {\bf 88}, 034017 (2013).



\bibitem{Cleven:2013sq} 
  M.~Cleven, Q.~Wang, F.~-K.~Guo, C.~Hanhart, U.~-G.~Meissner and Q.~Zhao,
  arXiv:1301.6461 [hep-ph].


 
\bibitem{Voloshin:2011qa} 
  M.~B.~Voloshin,
  {\it Phys.\ Rev.\ D} {\bf 84}, 031502 (2011).
  
  
\bibitem{Adachi:2012cx} 
  I.~Adachi {\it et al.}  [Belle Collaboration],
  arXiv:1209.6450 [hep-ex].
 
\end{thebibliography}
\end{document}